\documentclass[journal,final]{IEEEtran}
\usepackage{cite}
\usepackage{hyperref}
\usepackage[alsoload=synchem]{siunitx}
\DeclareSIUnit\pixel{pixel}
\ifCLASSINFOpdf
  \usepackage[pdftex]{graphicx} 
\else
  \usepackage[dvips]{graphicx}
\fi

\usepackage{tikz}

\newcommand\copyrighttext{
  \footnotesize \textcopyright 2016 IEEE. Personal use of this material is permitted.  Permission from IEEE must be obtained for all other uses, in any current or future media, including reprinting/republishing this material for advertising or promotional  purposes, creating new collective works, for resale or redistribution to servers or lists, or reuse of any copyrighted component of this work in other works.
  DOI: \href{https://dx.doi.org/10.1109/TNS.2016.2567448}{10.1109/TNS.2016.2567448}}
\newcommand\copyrightnotice{
\begin{tikzpicture}[remember picture,overlay]
\node[anchor=south] at (current page.south) {\fbox{\parbox{\dimexpr\textwidth-\fboxsep-\fboxrule\relax}{\copyrighttext}}};
\end{tikzpicture}
}

\usepackage[cmex10]{amsmath}

\usepackage[caption=false,font=footnotesize]{subfig}

\begin{document}

\title{Detector Damage at \\ X-Ray Free-Electron Laser Sources}

\author{G.~Blaj,\textsuperscript{*}\IEEEmembership{~Member,~IEEE},
        G.~Carini,\IEEEmembership{~Member,~IEEE},
        S.~Carron,
        G.~Haller,
        P.~Hart,
        J.~Hasi,
        S.~Herrmann,
        C.~Kenney,\IEEEmembership{~Member,~IEEE},
        J.~Segal,\IEEEmembership{~Member,~IEEE},
        C.A.~Stan,
        A.~Tomada,\IEEEmembership{~Member,~IEEE}
        
\thanks{Manuscript received December~21,~2015; revised March~27,~2016 and May~6,~2016. SLAC-PUB-16524.}
\thanks{G.~Blaj, G.~Carini, S.~Carron, G.~Haller, P.~Hart, J.~Hasi, S.~Herrmann, C.~Kenney, J.~Segal, C.A.~Stan and A.~Tomada are with SLAC National Accelerator Laboratory, Menlo Park, CA 94025, U.S.A.}
\thanks{\textsuperscript{*} Corresponding author: blaj@slac.stanford.edu.}
}

\markboth{}
{Shell \MakeLowercase{\textit{et al.}}: Bare Demo of IEEEtran.cls for Journals}

\maketitle
\copyrightnotice

\begin{abstract}

Free-electron lasers (FELs) opened a new window on imaging the motion of atoms and molecules. At SLAC, FEL experiments are performed at LCLS using 120~Hz pulses with 10\textsuperscript{12}~to~10\textsuperscript{13}~photons in 10~fs (billions of times brighter than at the most powerful synchrotrons). Concurrently, users and staff operate under high pressure due to flexible and often rapidly changing setups and low tolerance for system malfunction. This extreme detection environment raises unique challenges, from obvious to surprising, and leads to treating detectors as consumables. We discuss in detail the detector damage mechanisms observed in 7~years of operation at LCLS, together with the corresponding damage mitigation strategies and their effectiveness. Main types of damage mechanisms already identified include: (1)~x-ray radiation damage (from "catastrophic" to "classical"), (2)~direct and indirect damage caused by optical lasers, (3)~sample induced damage, (4)~vacuum related damage, (5)~high-pressure environment. In total, 19 damage mechanisms have been identified. We also present general strategies for reducing damage risk or minimizing the impact of detector damage on the science program. These include availability of replacement parts and skilled operators and also careful planning, incident investigation resulting in updated designs, procedures and operator training.

\end{abstract}



\section{Introduction}
Free-electron lasers (FELs) made it possible to image atoms and molecules at  time scales (\SI{10}{\femto\second}) and length scales (wavelengths of \SIrange[range-phrase= --]{1.3}{46}{\angstrom} corresponding to energies of \SIrange[range-phrase= --]{270}{9500}{\electronvolt}) relevant for atomic and molecular motion. This is achieved through intense x-ray laser pulses (\numrange[range-phrase = --]{e12}{e13}~photons, many orders of magnitude higher than at existing synchrotrons) with high coherence and short duration.

The first hard x-ray FEL, the Linac Coherent Light Source (LCLS), entered operation in 2009\cite{emma2010first} at the SLAC National Accelerator Laboratory. Other FELs are following: SACLA (first light in 2011)\cite{pile2011x}, the European XFEL, Swiss FEL, and PAL FEL.

X-ray detection is one of the essential components of x-ray synchrotron and FEL experiments. Synchrotrons have demanding detection requirements\cite{rajendran2011radiation}. However, FELs present a particularly challenging environment for x-ray detectors\cite{graafsma2009requirements,blaj2014detectors,blaj2014detector,blaj2015xray}, raising unique challenges, from obvious (e.g., radiation damage) to surprising (e.g., electromagnetic pulse damage -- EMP).

We will discuss the detectors in use at LCLS, the damage mechanisms observed, and corresponding damage mitigation strategies developed by the SLAC detector experts to minimize the chance of recurrence.

Previous works described radiation damage in semiconductors\cite{lutz1999semiconductor}, ASICs\cite{white2001radiation}, sensors\cite{wunstorf2001radiation}, and hybrid pixel detectors\cite{rajendran2011radiation} or summarized the experience of operating particular detectors at LCLS for short periods, e.g., pnCCD\cite{weidenspointner2011practical} and CSPAD\cite{tomada2012high}.

This article is the first work detailing all types of damage mechanisms observed in \num{7}~years of operation at LCLS as well as strategies to minimize chances of damage at FELs. This is particularly relevant for the growing FEL detection community (e.g., \cite{blaj2015xray,koch2013detector,graafsma2015integrating,blaj2015future}) and for other facilities operating detectors in challenging environments.

\section{Detection at LCLS}
Currently there are 6 instruments (or beamlines) at LCLS: AMO, SXR, XPP, XCS, CXI, and MEC. Typically only one of the 6 LCSL instruments is performing experiments at a time. If one of many essential experiment components would malfunction, the entire facility would temporarily be offline, unlike at synchrotrons. This results in a very low tolerance for instrument malfunction. The typical beamtime structure (5 consecutive days with \SI{12}{\hour} shifts) also requires demanding schedules for beamline operation and reconfiguration.

The detectors used most often at LCLS\cite{blaj2015xray} are CSPADs\cite{koerner2009x,philipp2011pixel,hart2012cornell} for hard x-rays and pnCCDs\cite{hartmann2008large} for tender x-rays (the line between "hard x-ray" and "tender x-ray" depends on the context; here we draw the line at \SI{5}{\kilo\electronvolt}), with other detectors used as needed (e.g., Rayonix, fCCD, Princeton, Hamamatsu etc.).

\subsection{CSPAD}

\begin{figure*}[!t]
\centerline{
\subfloat[]{\includegraphics[width=1.7in,height=1.3in]{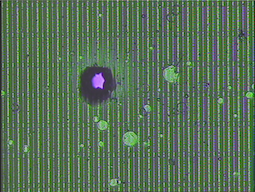}
\label{fig1a}}
\hfil
\subfloat[]{\includegraphics[width=1.7in,height=1.3in]{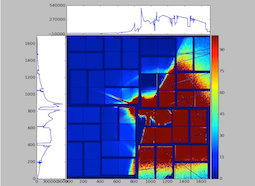}
\label{fig1b}}
\hfil
\subfloat[]{\includegraphics[width=1.7in,height=1.3in]{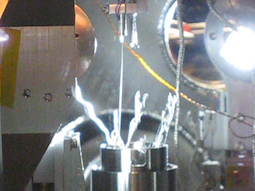}
\label{fig1c}}
\hfil
\subfloat[]{\includegraphics[width=1.7in,height=1.3in]{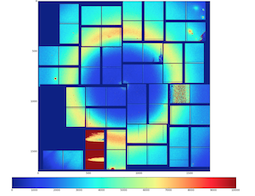}
\label{fig1d}}}
\caption{Catastrophic FEL x-ray beam damage:
(a) pnCCD plane hit by the focused direct beam in the CAMP chamber (reproduced with permission from \cite{weidenspointner2011practical});
(b) CSPAD~\SI{2.3}{\mega\pixel} camera hit by scattered radiation caused by the beam accidentally hitting a setup part \cite{tomada2012high};
(c) spontaneous formation of water crystals. Top: sample injector; bottom: catcher with crystals. Jet and crystals visible. Multiple crystals can be produced every second;
(d) water crystals in the direct beam produced intense diffraction spots which, in this instance, quasi-instantly damaged at least 3 ASICs of a CSPAD~\SI{2.3}{\mega\pixel} camera \cite{tomada2012high}.}
\label{fig1}
\end{figure*}

The CSPAD detectors\cite{koerner2009x,philipp2011pixel,hart2012cornell} are hybrid pixel detectors, typically using \SI{500}{\micro\metre} silicon sensors flip-chip bonded onto \SI{700}{\micro\metre} silicon read-out ASICs.

For hard x-rays there are 4 CSPAD~\SI{2.3}{\mega\pixel} cameras\cite{hart2012cspad} (2 at CXI, 1 at XPP, and 1 back-up). Three \SI{560}{\kilo\pixel} cameras\cite{herrmann2013cspad} are deployed at MEC. A constellation of about 16 small CSPAD~\SI{140}{\kilo\pixel} cameras\cite{herrmann2012cspad} are frequently redeployed as needed.

\subsection{pnCCD}
Each pnCCD camera is built around 2 silicon sensors with a thickness of about \SI{450}{\micro\metre}\cite{struder2010large} (called "half-planes").

For tender x-ray experiments, 2~pnCCD~\SI{1}{\mega\pixel} cameras are used, either together or independently. The pnCCD cameras were initially designed to be stationed at AMO in the CAMP chamber\cite{struder2010large} and its successor, the LAMP chamber\cite{carron2014lamp}. However, due to high demand, the pnCCD detectors are now traveling frequently between AMO, SXR, XCS and CXI.

\subsection{Other Detectors}
Other detectors with complementary characteristics to the CSPAD and pnCCD are used as necessary\cite{blaj2015xray}. They include fCCD camera (prototypes) and a custom Rayonix camera with 4 CCDs, a custom scintillator and taper with central beam hole.

\section{Catastrophic X-ray Damage}

\begin{figure*}[!t]
\centerline{
\subfloat[]{\includegraphics[width=1.7in,height=1.7in]{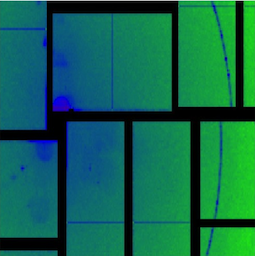}
\label{fig2a}}
\hfil
\subfloat[]{\includegraphics[width=1.7in,height=1.7in]{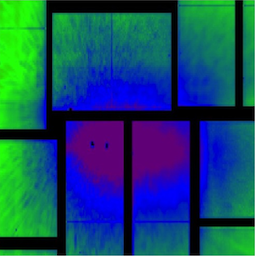}
\label{fig2b}}
\hfil
\subfloat[]{\includegraphics[width=1.7in,height=1.7in]{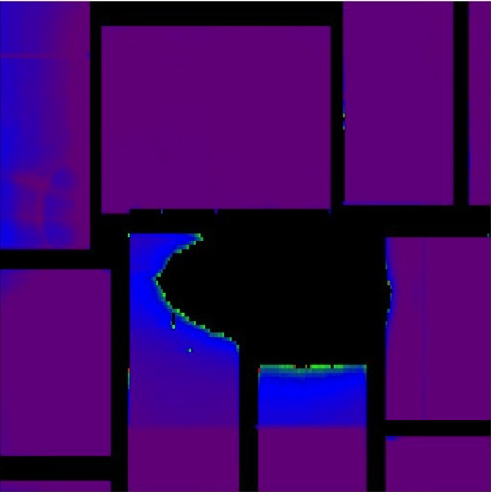}
\label{fig2c}}
\hfil
\subfloat[]{\includegraphics[width=1.7in,height=1.7in]{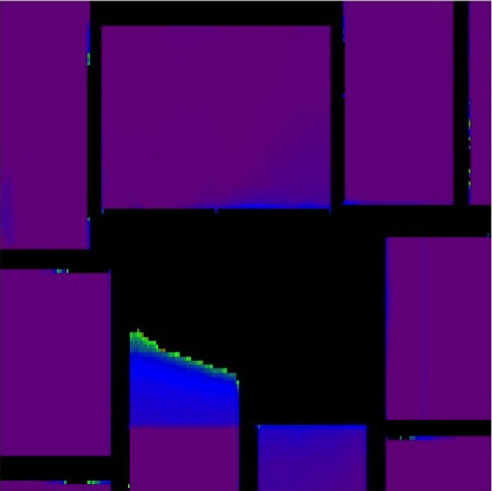}
\label{fig2d}}
}
\centerline{
\subfloat[]{\includegraphics[width=1.7in,height=1.7in]{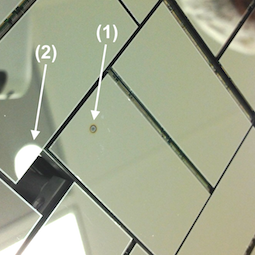}
\label{fig2e}}
\hfil
\subfloat[]{\includegraphics[width=1.7in,height=1.7in]{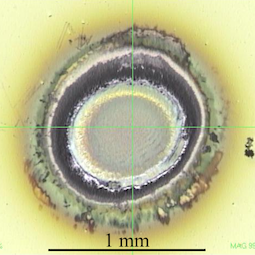}
\label{fig2f}}
\hfil
\subfloat[]{\includegraphics[width=1.7in,height=1.7in]{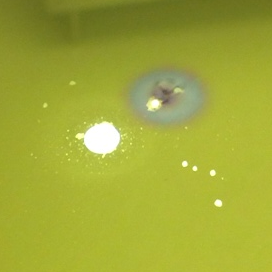}
\label{fig2g}}
\hfil
\subfloat[]{\includegraphics[width=1.7in,height=1.7in]{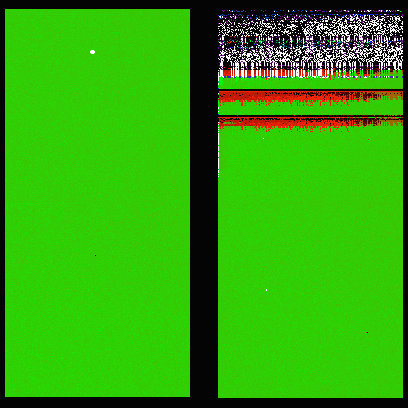}
\label{fig2h}}
}
\caption{Optical laser damage incident involving a CSPAD~\SI{2.3}{\mega\pixel} camera:
(a) last successful shot, showing diffraction pattern (geometric correction not applied);
(b) first optical laser shot into detector, high intensity pattern visible, all ASICs read out normally;
(c), (d) second and third optical laser shots into detector, one ASIC already damaged;
(e) resulting sensor damage by laser ablation (small dark spot), indicated by arrow (1); also semicircular surface contamination is visible (white), indicated by arrow (2);
(f) microscope close-up of damaged spot on the sensor (\SI{1}{\milli\metre} diameter, \SI{10}{\micro\metre} depth);
(g) damaged front shield (aluminum coated polyimide). Visible: central aperture (large bright spot), laser damage (small bright spots), diffuse laser reflection damage (discolored area);
(h) dark frame of the affected CSPAD module: focused beam damage (bright spot, top left) and diffuse scattering damage (ASIC, top right).}
\label{fig2}
\end{figure*}

We use the term "catastrophic x-ray damage" to emphasize the quasi-instantaneous character of the damage produced by high intensity x-ray pulses and distinguish it from the "classical", linear x-ray damage caused by individual photons\cite{lutz1999semiconductor} (typical for conventional radiation sources).

The catastrophic damage is similar to damage due to femtosecond laser micromachining\cite{perry1999ultrashort,gattass2008femtosecond}. The actual damage mechanisms include (depending on pulse duration and energy density): local melting, boiling, plasma formation, thermal shock resulting in cracking, and/or material removal\cite{perry1999ultrashort}.

Catastrophic x-ray radiation damage is a constant threat at FELs, due to accidental exposure to insufficiently attenuated beam, focused beam and formation of crystals reflecting the beam onto the detector. This can result in holes, damaged ASICs, or permanent damage spots.

\subsection{Focused Direct Beam}
The full direct beam can contain \numrange[range-phrase = --]{e12}{e13} photons per pulse at energies in the \si{\kilo\electronvolt} range, in pulses lasting femtoseconds. They can be focused down to micron size spots, yielding a very high energy density.

If the direct beam focused to, e.g., less than \SI{100}{\micro\metre} spots accidentally hits a detector, it can perforate the detector quasi-instantaneously. Such an incident happened during the first user experiment with the pnCCD cameras in the CAMP chamber\cite{weidenspointner2011practical}, see Fig.~\ref{fig1a}. Recurrence has been avoided in the subsequent \num{6}~years of LCLS operation.

\subsection{Unfocused Direct Beam or Scattered Beam}
Accidental reconfiguration of upstream x-ray optics (monochromators, attenuators, etc.), or the direct beam accidentally hitting setup parts can produce very intense scattering patterns over large areas of the camera.

While the energy density can be orders of magnitude lower than in the focused direct beam, the intensity is still high (orders of magnitude higher that at synchrotrons\cite{tomada2012high}) and has the potential to damage entire cameras within a few FEL pulses.

Fig.~\ref{fig1b} shows an incident with the FEL beam ($\sim$~\SI{8}{\kilo\electronvolt}) accidentally hitting the ceramic support of a repeller-extractor structure (part of a time of flight spectrometer) and projecting an intense scattering pattern onto half of a CSPAD~\SI{2.3}{\mega\pixel} camera. After the incident, the central areas of the camera had significantly elevated dark and noise levels\cite{tomada2012high}.

\subsection{Diffraction on Ice Icicles}
Aqueous jets injected in vacuum are often used to deliver the sample. At vacuum chamber pressures from tens to hundreds of \si{\milli\torr}, these jets freeze upon impact onto the sample-catching vessel and lead to the growth of multiple ice icicles, e.g., Fig.~\ref{fig1c}.

When the direct beam hits one of these icicles, it projects intense diffraction patterns, resulting into fluxes per pixel of, e.g., \numrange{e10}{e11}~photons per \SI{100}{\femto\second} pulse (\numrange[range-phrase = --]{e12}{e14} higher than in a typical synchrotron\cite{tomada2012high}), which can quickly damage cameras.

Fig.~\ref{fig1d} shows the aftermath of such an incident on a CSPAD~\SI{2.3}{\mega\pixel} camera. After this incident, at least 3 ASICs were permanently damaged\cite{tomada2012high}. Currently the operators are aware of this risk, monitoring and adjusting jet parameters to avoid icicle formation. This is one of the leading causes of spot damage (clusters of several pixels).

\subsection{Mitigation}
We avoid projecting high intensity signals on the cameras by inserting all attenuators before changing configurations, to minimize the impact of accidental high intensities. Then the attenuation is gradually reduced while monitoring the detected signal, until reaching the desired level.

Hybrid pixel detectors, including the CSPAD, typically have large ASIC areas with sensitive pixels, the "pixel matrix", covered by the sensor. In addition, they have one small area of the ASICs protruding from underneath the sensors, the "ASIC balcony".

Sensor or ASIC damage in the pixel matrix is often localized, disabling spots, columns and/or rows on detectors. However, the ASIC balcony is one of the most sensitive areas to radiation damage (both catastrophic and classical), with the potential of disabling entire ASICs.

Radiation damage of entire ASICs was greatly reduced by using high-Z shields over the exposed ASIC balcony\cite{tomada2012high}. However, local damage can't be prevented in high intensity x-ray imaging. If damage occurs, annealing only slightly improves the damaged areas\cite{tomada2012high}.

The pnCCD cameras are more sensitive to radiation damage, less easy to modify and repair, and significantly more expensive and difficult to acquire than the CSPAD cameras. Because engineering solutions are not feasible, we compensate by strictly evaluating the risks, including radiation damage, before each experiment. If the risk is deemed high, detector experts participate in the experiments and monitor the camera for the entire duration of the experiment.

Once radiation damage occurs, its effects can be corrected to some extent by updating the dark maps and masking the pixels that become bright or noisy.

While catastrophic radiation damage can't be eliminated, its impact can be minimized effectively through (1) using a modular design which facilitates the exchange of damaged modules and (2) having spare detector modules available for replacement. On average, we are replacing $\sim$~one CSPAD detector module per month and $\sim$~two pnCCD half-planes per year.

\section{Gradual X-ray Radiation Damage}
This is the "classical" radiation damage induced by typical radiation sources, described in literature on semiconductors\cite{lutz1999semiconductor}, ASICs\cite{white2001radiation}, sensors\cite{wunstorf2001radiation}, and hybrid pixel detectors\cite{rajendran2011radiation}.

Radiation damage manifests through increased sensor leakage and ASIC damage in the exposed areas. These lead to elevated dark currents and noise, respectively.

\subsection{Water Rings}
Water jets are often used for sample delivery. The x-ray beam interacting with water produces water scattering rings similar to the ones shown in Fig.~\ref{fig1d}. Often the useful signal is collected in the same areas\cite{sellberg2014ultrafast}, making it impossible to shield the detectors with filters or attenuators. As a result, radiation damage is slowly building up in the exposed areas.

\subsection{Diffraction Spots}

Sometimes the sample can project intense diffraction spots onto the camera. While the "Diffraction on Ice Icicles" subsection above also refers to damage caused by diffraction spots, we artificially include a subset of "Diffraction on Ice Icicles" events in this category based on the effects: disabling small clusters of pixels, in contrast to disabling entire ASICs.

For example, in the case of aqueous jets, diffraction much more intense than expected from the sample can occur if liquid jets malfunction and freeze, or if the sample liquor contains strongly diffracting crystals of compounds used in the preparation of sample, e.g., ammonium sulfate.

Although damage effects in this category are usually limited, a high quality crystal can scatter a significant fraction of the full beam. This could lead to, e.g., a diffraction spot with \num{e+11}~photons impacting a \SI{10x10}{\micro\metre\square} area of the detector. In hard x-ray experiments (often using \SI{8}{\kilo\electronvolt} photons and CSPAD detectors), such a diffraction spot would result in a dose of $\sim$~\SI{60}{\gray} per \SI{100}{\femto\second} pulse in the impacted ASIC oxide gates. This dose is calculated taking into account the shielding provided by the \SI{500}{\micro\metre} silicon sensor.

In a best case scenario, only seconds might elapse until such a condition is noticed and corrected, resulting - at \SI{120}{\hertz} - in doses (likely far) exceeding
\SIrange{50}{100}{\kilo\gray} and impacting the same ASIC area in a short time. This is comparable with the dose that might be delivered to the ATLAS innermost layer, the ATLAS Pixel Detector, in years\cite{tomada2012high}. This dose is also delivered in extremely short and intense pulses of \SIrange{10}{100}{\femto\second}.

\subsection{Mitigation}
The impact of the gradual radiation damage is usually low: detector areas with somewhat elevated noise or several small clusters of disabled pixels usually perform well. The effects can be mitigated with updated dark frames and/or bad pixel maps. Consequently, we do not include gradual radiation damage in the risk analysis in section~X.

The mitigation measures mentioned above (section III D) also apply to radiation damage. Additionally, accidental radiation damage is further reduced at some of the instruments by actively monitoring the amount of radiation on the camera and quickly deflecting the beam if the intensity is higher than expected.

\section{High Power Optical Lasers}

High power optical lasers are used for pumping samples and for high-energy density science (HED) experiments. They can damage detectors or impede data acquisition when the beam impacts the detector or when the sample interaction takes place close to the detector, respectively.

\subsection{Direct Beam}
The beam of optical lasers accidentally hitting detectors can have dramatic effects\cite{perry1999ultrashort,gattass2008femtosecond}, similar to the catastrophic x-ray damage mentioned in section III.

This occurred in an experiment where the optical pump laser had to be directed towards the detector (a CSPAD~\SI{2.3}{\mega\pixel} camera). A beam stop was placed. After reconfiguration during the experiment, the laser beam missed the beam stop and impacted the detector instead (Fig.~\ref{fig2}).

Several focused pulses of the optical laser impacted the camera (see Fig.~\ref{fig2a}--\ref{fig2d}), with the direct beam drilling a hole into the sensor (Fig.~\ref{fig2e}, \ref{fig2f}) and the diffuse scattering damaging the neighboring ASIC (Fig.~\ref{fig2h}).

\subsection{Electromagnetic Pulse}
High power optical lasers are often used (e.g., $\SI{1}{\joule}/\SI{40}{\femto\second}=\SI{25}{\tera\watt}$ in high energy density (HED) experiments, interacting centimeters away from the detectors). This increases the risk of damage or impedes data acquisition through electromagnetic pulses (EMP)\cite{eder2010assessment}. Initial tests with CSPAD~\SI{140}{\kilo\pixel} cameras resulted in damaged electronics and pulses traveling along cables and causing malfunction of the downstream DAQ servers.

\subsection{Mitigation}
High power optical lasers should ideally be pointed away from detectors, focused outside the detector plane, and shielded with beam stops. We found that only one of these precautions is not sufficient, as the chamber layout is often optimized during experiments. Unfortunately, existing experimental chamber constraints and changing setups do not always allow implementing all these precautions.

\begin{figure}[!t]
\includegraphics[width=\columnwidth]{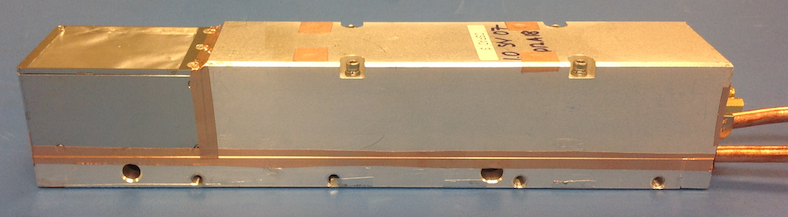}
\caption{Careful EMP shielding (copper tape) and electrically conducting front shield (aluminized black polyimide, top left) made it possible to operate cameras within centimeters to the interaction point in HED experiments; CSPAD~\SI{140}{\kilo\pixel} camera shown.}
\label{fig3}
\end{figure}

The EMP shielding and grounding were gradually improved for the relevant CSPAD~\SI{560}{\kilo\pixel} and \SI{140}{\kilo\pixel} cameras and cables (e.g., Fig.~\ref{fig3}) by carefully shielding the cameras and cables with vacuum compatible copper tape and aluminum foils, respectively. Improved shielding eliminated detector damage, however, some frames were still corrupted. Further shielding refinement eliminated data acquisition errors and led to reliable operation currently.

\section{Sample Damage}

\begin{figure*}[!t]
\centerline{
\subfloat[]{\includegraphics[width=1.7in,height=2.28in]{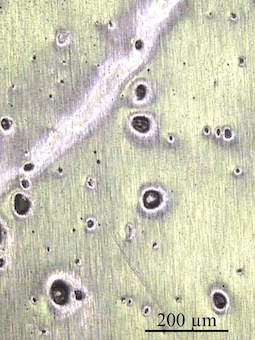}
\label{fig4a}}
\hfil
\subfloat[]{\includegraphics[width=1.7in,height=2.28in]{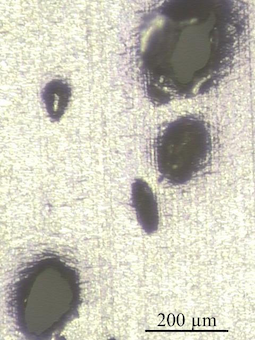}
\label{fig4b}}
\hfil
\subfloat[]{\includegraphics[width=1.7in,height=2.28in]{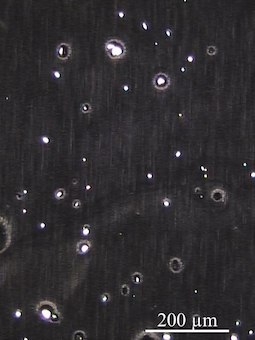}
\label{fig4c}}
\hfil
\subfloat[]{\includegraphics[width=1.7in,height=2.28in]{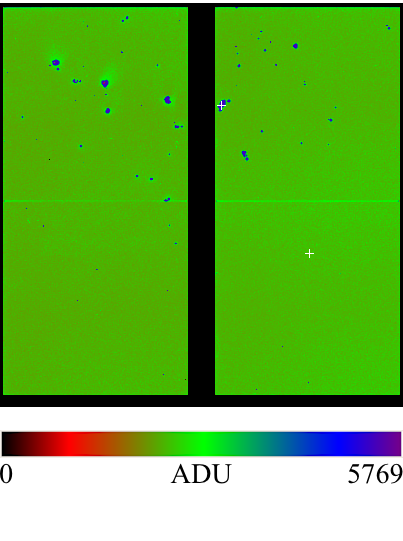}
\label{fig4d}}}
\caption{Shrapnel damage with different shields in high energy density experiments: (a) aluminum shield; (b) aluminized polyimide shield; (c) aluminized black polyimide shield; (d) Damaged sensor in CSPAD~\SI{140}{\kilo\pixel} camera, dark frame.}
\label{fig4}
\end{figure*}

The sample can contaminate detector surfaces (and other components), or it can produce shrapnel. Contamination can reduce quantum efficiency with softer x-rays or damage the sensors.

\subsection{Sensor Contamination}
The sample (solid or liquid) interacting with the FEL beam is often accelerated and deposited on surfaces. While most of the sample that does not interact with the FEL beam is caught by the catcher, a fraction can diffuse in the chamber and get deposited on surfaces, including detectors (e.g., Fig.~\ref{fig2e}, bright half circle up and to the left of the central beam opening). The film on the detector surface can reduce quantum efficiency at low photon energies or even damage the sensor.

The direct beam can also accidentally hit parts of the setup. The resulting debris can contaminate the detectors\cite{weidenspointner2011practical}, having a similar effects to sample contamination.

\subsection{Shrapnel}

High energy deposition in small areas can cause (parts of) the sample to explode, expelling high velocity pieces of shrapnel that can impact the detector\cite{eder2010assessment}. This is especially true in HED type experiments, where the detectors can be within centimeters to the sample.

Fig.~\ref{fig4a}--\ref{fig4c} shows examples of shrapnel damage on different shields and Fig.~\ref{fig4d} shows an example of sensor damage caused by nano- and micro-shrapnel.

\subsection{Sensor Etching}
\begin{figure}[!t]
\centerline{
\subfloat[]{\includegraphics[width=1.7in,height=1.7in]{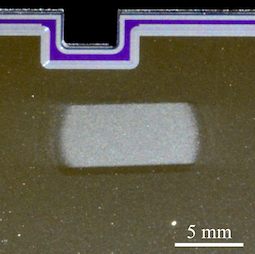}
\label{fig5a}}
\hfil
\subfloat[]{\includegraphics[width=1.7in,height=1.7in]{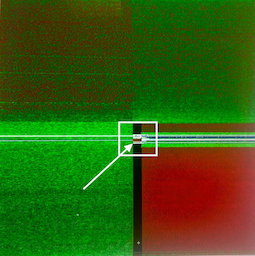}
\label{fig5b}}
}
\caption{Accidental xenon sputter etching of a pnCCD sensor: (a) shows a close-up of the affected rectangular area of a half-plane, surrounded by the intact sensor surface. The inner edge and central beam hole are visible at the top; (b) shows the leakage and noise of a full plane. The white box indicates the edges of close-up in Fig.~\ref{fig5a}. The xenon etching resulted in a significant increase in leakage and noise, disabling the damaged area and corresponding rows and columns. The arrow shows the location of damage.}
\label{fig5}
\end{figure}

High velocity sample clusters impacting sensor surfaces can etch them, as it was accidentally observed in a pnCCD experiment with xenon clusters. It was deemed necessary from a scientific point of view to operate without a detector shield. The xenon clusters formed were sufficiently energetic to etch the pnCCD sensor surface, damaging the thin aluminum coating and the silicon crystal structure (Fig.~\ref{fig5a}) and leading to a dramatic increase in the noise level in the affected area and corresponding rows and columns (Fig.~\ref{fig5b}).

\subsection{Mitigation}
To reduce the damage rate while fulfilling the different requirements of individual experiments (i.e., optical shielding, x-ray transparency for used photon energies), various front shields have been produced. These are manufactured from different materials with a range of thicknesses: black polyimide, aluminum-coated polyimide, aluminum-coated black polyimide, polyethylene, aluminum and beryllium, often with a central beam pipe for the large cameras.

These shields greatly reduce detector surface contamination, however, they do not eliminate it completely. They also protect the detectors from spurious signals induced by stray light. However, they are not very effective against shrapnel and need to be replaced regularly.

In HED experiments, the x-ray pattern intensity is often high enough to allow using thick polyethylene shields (\SI{125}{\micro\metre}) stacked on top of the shields. These greatly reduce the shrapnel damage, with a limited impact on quantum efficiency (e.g., \SI{95}{\percent} at \SI{10}{\kilo\electronvolt} and \SI{60}{\percent} at \SI{5}{\kilo\electronvolt}).

\section{Vacuum}
Many CSPAD and all pnCCCD cameras are operating in experimental chambers with ultra high vacuum (UHV), presenting strict design, surface contamination and cooling requirements.

\subsection{Cooling Failure}
CSPAD detector modules or entire cameras occasionally lose thermal contact with the cooling circuit ($\sim$~twice a year), resulting in rapid overheating. In vacuum, this can quickly destroy cameras. Using temperature and sensor bias current interlocks prevents damage and alerts operators.

The cooling of pnCCD cameras is particularly complex\cite{carron2014lamp}. The heat and the mechanical stresses due to (1) rapid temperature changes or (2) cooling outside the safe temperature range could damage pnCCD cameras. To prevent damage, the sensor temperature  is constantly monitored during experiments (operating at about \SIrange{-70}{-60}{\celsius}) and is set to safer values between experiments (about \SIrange{-50}{-40}{\celsius}).

\subsection{Explosive (De)Compression}
Ideally, pnCCD cameras should always be under vacuum, whether operating or not. Many CSPAD cameras are also under vacuum during experiments. Occasionally, the vacuum chambers housing the detectors need to be opened. Venting should be gentle, with the detector temperature close to room temperature.

In one incident, the chamber housing 2 pnCCD cameras at $\sim$~\SI{-45}{\celsius} was quickly vented, damaging one half plane.

While such incidents can be prevented by locking and tagging the vacuum chamber when the detectors are unprotected inside, this approach would also hinder essential work.

\subsection{Electrostatic Discharge Between Detector Wire-Bonds}
In CSPAD cameras, wire-bonds connect the ASICs and sensors to the underlying carrier board. One of these wire-bonds provides a \SI{200}{\volt} bias to the typical silicon sensor and is typically located $\sim$\SI{500}{\micro\metre} from neighboring wire-bonds. However, accidental damage might bend the bias wire-bond or neighbors within microns (or even short them, see section VIII B below). While at atmospheric pressure and in vacuum the breakdown voltage is higher than the bias voltage, particular combinations of gases and pressures can cause arcing between these wire-bonds (Paschen's Law).

In an experiment using CSPAD cameras, an accidental increase in the He pressure in the vacuum chamber was correlated with repeated tripping of the sensor bias current interlock and eventual failure in one CSPAD module. It was later determined that arcing between the bias wire-bonds and neighboring wire-bonds caused this incident.

Unintentional pressure spikes due to sample delivery are prevented now through updated sample delivery software.

\subsection{Electrostatic Discharge from Sample Injector}
One of the types of liquid sample injectors used in serial crystallography experiments\cite{sierra2012nanoflow} uses large electric fields to pull (or electrospin) liquid jets with diameters of a few microns from silica capillaries with inner diameters between \SIrange[range-phrase=~and~]{50}{100}{\micro\metre}. The electric fields are applied between the tip of the capillary and a centimeter-sized counter-electrode located approximately \SI{1}{\centi\metre} from the tip.

In normal operation, the injector setup uses electrical potentials between \SIrange[range-phrase=~and~]{5}{10}{\kilo\volt}. As the gap between the tip and the counter-electrode is smaller than the separation between the injector and the vacuum chamber components (including the CSPAD), electrical discharges typically occur within the injector setup.

In one LCLS beamtime, this type of injector was part of a complex setup including optical mirrors in anodized aluminum mounts, which due to the oxide layer may not provide proper electrical grounding. During this experiment, the counter-electrode was set at an electrical potential of \SI{-6}{\kilo\volt} to compensate for the presence of air bubbles in the capillary. These bubbles opened the electrical circuit formed by the electrically conductive liquid sample and prevented the sample injection. Under these operating conditions, the CSPAD camera suffered a failure that was later diagnosed as electrostatic discharge from the sample injector.

We hypothesize that once the electrical circuit formed by the liquid was interrupted by bubbles, the optical mounts located within \SIrange{1}{2}{\centi\metre} of the \SI{-6}{\kilo\volt} counter-electrode charged electrically because of imperfect grounding. As these mounts reached to within a few centimeters of the CSPAD camera assembly, we hypothesize that an electrical discharge occurred between the counter-electrode and the CSPAD via the optical mounts. We note that due to the evaporation of the liquid sample, the pressure in the vacuum chamber can routinely reach a few milliTorrs, increasing the possibility of electrical breakdown.

After this failure, the counter-electrode was always operated grounded, and no further failures occurred.

\section{Handling - Time Pressure}
Because typically only one experiment is active at a time, there is a high pressure for each experiment to be successful. 

Both during and between experiments it is often necessary to reconfigure the beamlines for the next (phases of) experiments. Combined with the usually ambitious setups for most experiments, this results in intense activity at the beamline both during and between experiments. This high pressure environment sometimes results in detector damage.

In addition, detectors have to be moved more or less frequently to different locations within the experimental setup, or occasionally to other instruments. This increases the risk of damage.

In particular, the pnCCD cameras were initially designed to be stationed at AMO in the CAMP chamber\cite{struder2010large} and its successor, the LAMP chamber\cite{carron2014lamp}. However, due to high demand, the pnCCD detectors are now traveling frequently between AMO, SXR, XCS and CXI.

\subsection{Collisions with Detector}
\begin{figure*}[!t]
\centerline{
\subfloat[]{\includegraphics[width=1.7in,height=1.7in]{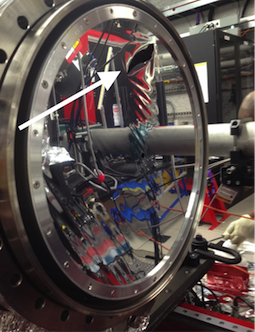}
\label{fig6a}}
\hfil
\subfloat[]{\includegraphics[width=1.7in,height=1.7in]{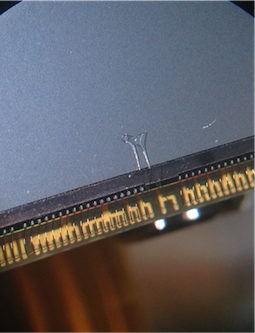}
\label{fig6b}}
\hfil
\subfloat[]{\includegraphics[width=1.7in,height=1.7in]{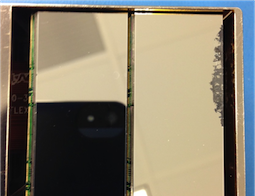}
\label{fig6c}}
\hfil
\subfloat[]{\includegraphics[width=1.7in,height=1.7in]{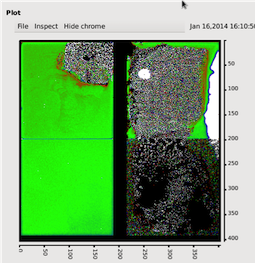}
\label{fig6d}}}
\caption{Handling damage: (a) detector collision resulting in damaged shield on CSPAD~\SI{2.3}{\mega\pixel} camera, with damaged area indicated by the arrow; (b) damaged wire-bonds; (c) water damage on CSPAD~\SI{140}{\kilo\pixel}; (d) corresponding dark frame.}
\label{fig6}
\end{figure*}

Experimental chambers are complex, with tight limits and flexible layouts that need to be changed from experiment to experiment or even during  experiments. The high degree of flexibility required by the experiments limit the implementation of engineering controls. Consequently, accidental collisions and damaged front shields have been observed $\sim$~twice a year, e.g., Fig.~\ref{fig6a}.

For the short term, we maintain a supply of spare shields and replace them as necessary. For the long term, there is a tendency to move towards sets of pre-defined configurations for each instrument, which will simplify and standardize the support for experiments, facilitating implementation of engineering controls.

\subsection{Damaged Wire-Bonds}
Sometimes, exchanging detector shields or redeploying cameras resulted in wire-bond damage, e.g., Fig.~\ref{fig6b}. This risk was greatly reduced by having only detector experts perform particularly delicate handling, including shield replacements.

Another significant reduction of damage incidence was achieved by modifying the relevant camera mechanical assemblies to allow the operators to easily stack and exchange filters onto the relevant detector shields.

\subsection{Electrostatic Discharge through Cables}
Electrostatic discharge damage happens occasionally ($\sim$~\numrange{4}{6} times a year) while moving detectors and DAQ systems or connecting or disconnecting cables, resulting in damaged readout electronics or DAQ components. The incidence was greatly reduced by observing ESD precautions, documenting procedures, and having delicate moving and cabling tasks performed only by detector experts.

\subsection{Leaky Cooling Connections}
The CSPAD water cooling uses brazed copper pipes with Swagelok~VCR or Swagelok~Quick-Connect fittings; these pipes and fittings are delicate, requiring careful handling. During an experiment, a CSPAD~\SI{140}{\kilo\pixel} camera was converted from VCR to Quick-Connect to match the existing chiller fittings. The resulting connection was faulty, causing water leakage on the powered detector and resulting in detector damage (Fig.~\ref{fig6c},~\ref{fig6d}).

Experiments are now provided complete packages (detector, matching chillers, tubing, connectors) to eliminate the need of modifying connections. While cooling in-air detectors under the dew point would have a similar effect, operators carefully avoided such an incident.

\subsection{Mechanical Vibrations and Shocks}
To reduce vibration and shock damage, all CSPAD cameras are transported and stored in heavily padded cases (e.g., Pelican). The pnCCD cameras are moved on specially built carts with air filled casters. These also allow secure rotation of the cameras to orientations matching the different experimental chambers. A Rayonix camera suffered damage from mechanical vibrations during a move between instruments; a dedicated transport case for this camera is being prepared.

\section{Damage Prevention}

\begin{table*}
  \renewcommand{\arraystretch}{1.3}
  \caption{Overview of detector damage mechanisms, mitigation measures and their effectiveness}
  \label{table_mitigation}
  \centering
  \begin{tabular}{*{4}{l}}
    \hline
    Category & Mechanism & Mitigation & Effectiveness \\
    \hline
     & ~1. focused direct beam & high attenuation before changes,& no subsequent incidents \\
     & &  then gradual decrease & \\
    X-ray, catastrophic & ~2. unfocused direct beam & high Z strips & reduced ASIC damage rate\textsuperscript{*} \\
     & ~3. ice icicle diffraction & operators watch for crystal formation & few subsequent incidents \\
     & &  and adjust injection & \\     
    \hline
    X-ray, gradual & ~4. water rings & high Z strips & reduced ASIC damage rate \\
     & ~5. diffraction spots & high Z strips & reduced ASIC damage rate\textsuperscript{*} \\
    \hline
    Optical lasers & ~6. direct beam & --- (experiment, chamber restrictions) & --- \\
     & ~7. electromagnetic pulse & careful EMP shielding & no subsequent incidents \\
    \hline
     & ~8. sensor contamination & front shields & reduced sensor contamination \\
    Sample & ~9. shrapnel & thick (e.g., \SI{125}{\micro\metre}) polyethylene & thick polyethylene is effective, \\
    & 10. sensor etching & front shields & reduced sensor damage \\
     & & shields when possible & other shields not \\
    \hline
     & 11. cooling failure & temperature and bias current interlocks & no subsequent incidents \\
    Vacuum & 12. explosive (de)compression & --- & --- ("lock and tag" would \\
     & & &  hinder essential work) \\    
     & 13. discharge between wire-bonds & updated sample delivery software & no subsequent incidents \\
     & 14. discharge from sample injector & counter-electrode grounded & no subsequent incidents \\
    \hline
     & 15. collisions with detector & --- (availability of spare parts) & --- (beamlines often reconfigured) \\
    Handling & 16. damaged wire-bonds & certain operations: detector experts only & incidence almost eliminated \\
    (time pressure) & 17. electrostatic discharge & certain operations: detector experts only & significantly reduced incidence \\
     & 18. leaky cooling connections & certain operations: detector experts only & no subsequent incidents \\
     & 19. mechanical vibrations, shocks & dedicated transport carts and cases & no subsequent incidents \\
    \hline
    \multicolumn{4}{l}{\textsuperscript{*} while greatly reducing the ASIC damage rate, the high-Z strips do not prevent radiation damage in the pixel matrix} \\
  \end{tabular}
  \par
\end{table*}

\subsection{Planning}
It is impossible to eliminate risks altogether, thus the approach is to maximize science output while minimizing risks to current and subsequent experiments.

The most effective damage mitigation approach is using modular camera designs and maintaining availability of replacement modules and of experts skilled in camera maintenance.

Each experiment carries some risk of detector damage. When the risk is deemed to be high, detector experts, instrument scientists and users discuss risk minimization strategies that would have a minimal impact on the experiment.

The feasibility of repairing possible damage and the risk to subsequent experiments are also weighted. This leads to a decision on the detectors to use, their location, shielding, and contingency plans in case of detector damage.

The pnCCD is particularly sensitive to all types of damage. At the same time, because it is not an in-house detector, it is much more complex and expensive to maintain and repair and spare module availability is limited. For this reason, experiments with the pnCCD detectors are scrutinized much more closely for damage risks. Often detector experts will actively participate in experiments and closely monitor the cameras when the risk of damage is high.

\subsection{Specialization}
We found that specialization (the same detector experts performing the same tasks) allows these members to acquire a deeper knowledge of the particular details of each detector, potential issues, and better dexterity, leading to a greatly reduced rate of incidents.

In practice we aim to have (1) the same operators at beamlines and (2) particularly delicate tasks (e.g., exchanges of shields, replacement of modules, etc.) performed by the detector experts most experienced in that task.

\subsection{Incident investigation}
While some damage can't be prevented, every incident or near incident is promptly investigated by detector experts and yields valuable lessons for future damage prevention. We define "near incidents" as risky occurrences that do not result in detector damage, e.g., detector collisions, unauthorized modifications, etc. The conclusions are discussed with the detector operators, instrument scientists etc., leading to updated camera designs, operating procedures or engineering controls (when possible) or to increased awareness of potential problems and their mitigation. Disseminating the investigation results to operators greatly reduce incident recurrence.

See Table \ref{table_mitigation} for a summary of observed damage mechanisms, corresponding mitigation measures, and their effectiveness.

\section{Risk Analysis}

\subsection{CSPAD}

The damage rate depends on factors including the instrument, the type of experiment, type of camera, handling, etc. The influence of each factor is difficult to estimate due to limited statistics. However, aggregated results are statistically significant and relevant for similar facilities.

Table~\ref{table_damage} shows a quarterly overview of the number of installed CSPAD modules, average running time and damage rate over the most recent \SI{2.5}{years} period.
\begin{table}[!t]
  \renewcommand{\arraystretch}{1.3}
  \caption{Quarterly overview of CSPAD modules installed, running time, damaged modules and damage rate}
  \label{table_damage}
  \centering
  \begin{tabular}{{r}{l}*{4}{r}}
    \hline
    Nr. & Quarter & Installed  & Usage\textsuperscript{*} & Damaged & Damage rate \\
    & & modules & \si{\hour} & modules & \si{\per\hour} \\
    \hline
    1 & Q4 2013 & \num{67} & \num{10709} & \num{5} & \num{4.67E-4} \\
    2 & Q1 2014 & \num{72} & \num{10091} & \num{7} & \num{6.94E-4} \\
    3 & Q2 2014 & \num{72} & \num{12493} & \num{3} & \num{2.40E-4} \\
    4 & Q3 2014 & \num{72} & \num{18485} & \num{3} & \num{1.62E-4} \\
    5 & Q4 2014 & \num{71} & \num{18999} & \num{1} & \num{5.26E-5} \\
    6 & Q1 2015 & \num{71} & \num{16030} & \num{1} & \num{6.24E-4} \\
    7 & Q2 2015 & \num{70} & \num{19237} & \num{0} & \num{0} \\
    8 & Q3 2015 & \num{70} & \num{17141} & \num{2} & \num{1.17E-4} \\
    9 & Q4 2015 & \num{77} & \num{18420} & \num{5} & \num{2.71E-4} \\
    10 & Q1 2016 & \num{79} & \num{20223} & \num{4} & \num{1.98E-4} \\
    \hline
    & Total & & \num{161828} & 31 & \\
    \hline
    \multicolumn{6}{l}{\textsuperscript{*} Across all installed modules} \\
  \end{tabular}
\end{table}

For a total running time of \SI{161828}{\hour} with \num{31} failures, the mean time between failures for CSPAD modules is $MTBF=\SI{161828}{\hour}/\num{31}=\SI{5220}{\hour}$. Assuming an exponential model, the corresponding \SI{95}{\percent} confidence limits are $MTBF_{lower}=\SI{3676}{\hour}$ and $MTBF_{upper}=\SI{7689}{\hour}$ for individual modules and \SI{140}{\kilo\pixel} cameras. \SI{560}{\kilo\pixel} cameras use 4 modules and \SI{2.3}{\mega\pixel} cameras use 16 modules, reducing the $MTBF$ by a factor of 4 and 16, respectively.

We hypothesize that the failure rate is on a downward trend over this period. By analyzing the slope of a simple linear fit of the failure rate as a function of quarter number, we can test the null and the alternate hypothesis:
\begin{itemize}
\item $H_0$: slope smaller than zero (damage rate decreasing)
\item $H_a$: slope equal to zero (damage rate constant)
\end{itemize}
We obtained a slope of $\SI{4.860E-4}{h^{-1}}$ with a standard error $\SI{3.450E-4}{h^{-1}}$, corresponding to a t-score of \num{1.401}. For $10-2=8$ degrees of freedom, $P(t<1.401)=0.9006$, confirming the null hypothesis at \SI{90}{\percent} confidence level. It is thus likely that the failure rate continued to decrease over the period reported in Table~\ref{table_damage}.

\subsection{pnCCD}
There are 2 pnCCD cameras consisting of 2 modules each.
In 2014 the modules have been operated for a total of $\SI{1344}{\hour}$ with 2 damage incidents, resulting in a mean time between failures $MTBF=\SI{672}{\hour}$. Due to the limited statistics, the \SI{95}{\percent} confidence limits are $MTBF_{lower}=\SI{185}{\hour}$ and $MTBF_{upper}=\SI{5549}{\hour}$ for individual modules. For a camera consisting of 2 modules, the $MTBF$ is half of the numbers shown above.

\section{Conclusion}
At FELs, detectors are consumables. Free-electron lasers are a challenging environment for detectors: the combination of demanding technical requirements and flexibility requirements render some detector damage unavoidable.

The best mitigation of damage effects is to have spare modules available and use modular designs (which facilitate easy replacement). Having on-site experts who can maintain cameras make rapid interventions (in hours) possible.

The risk of damage can be reduced further by careful planning before experiments and encouraging detector experts to specialize in subsets of particularly delicate operations.

While some damage can't be prevented, (near) incidents yield valuable lessons which help decrease future risks.

\appendices

\section*{Acknowledgments}

Portions of this research were carried out at the Linac Coherent Light Source (LCLS) at the SLAC National Accelerator Laboratory. LCLS is a Office of Science User Facility operated for the U.S. Department of Energy Office of Science by Stanford University.

The authors would like to thank current and former members of the LCLS community for their contribution to this research: C.~Bostedt, S.~Boutet, M.J.~Bucher, P.~Caragiulo, J.C.~Castagna, D.S.~Damiani, G.~Dakovski, A.~Dragone, M.~Dubrovin, B.~Duda, T.~Ekeberg, D.~Fritz, A.~Fry, E.~Galtier, M.~Hayes, P.~Heimann, R.~Herbst, D.~Kiehl, J.~Koglin, H.J.~Lee, H.~Lemke, M.~Liang, L.~Manger, B.~Markovic, M.~McCulloch, M.~Messerschmidt, A.~Mitra, B.~Nagler, S.~Nelson, T.~Nieland, K.~Nishimura, S.~Osier, J.~Pines, L.~Salgado, L.~Sapolnikov, M.~L.~Swiggers, J.~Thayer, M.~Weaver, G.~Williams, X.~Zhou and D.~Zhu.

\ifCLASSOPTIONcaptionsoff
  \newpage
\fi

\bibliographystyle{IEEEtran}

\bibliography{IEEEabrv,blaj}

\end{document}